\documentclass[12pt]{article}
\title{
Spontaneous SU(2) symmetry violation in the $SU(2)_L \times SU(2)_R\times
SU(4)$ electroweak  model}
\author{Yu.A.Simonov\\ Institute of Theoretical and Experimental
Physics\\ 117118, Moscow, B.Cheremushkinskaya 25, Russia}
\date{}

\newcommand{\ga}{\gtrsim}

\newcommand{\beq}{\begin{eqnarray}}
 \newcommand{\eeq}{\end{eqnarray}}
\newcommand{\be}{\begin{equation}}
 \newcommand{\ee}{\end{equation}}

\def\ga{\mathrel{\mathpalette\fun >}}
\def\fun#1#2{\lower3.6pt\vbox{\baselineskip0pt\lineskip.9pt
\ialign{$\mathsurround=0pt#1\hfil ##\hfil$\crcr#2\crcr\sim\crcr}}}

\newcommand{{\SD}}{\rm SD}

\newcommand{\vex}{\mbox{\boldmath${\rm x}$}}
\newcommand{\vey}{\mbox{\boldmath${\rm y}$}}

\newcommand{\vesig}{\mbox{\boldmath${\rm \sigma}$}}

\newcommand{\veR}{\mbox{\boldmath${\rm R}$}}

\newcommand{\ven}{\mbox{\boldmath${\rm n}$}}

\newcommand{\vexi}{\mbox{\boldmath${\rm \xi}$}}

\newcommand{\veH}{\mbox{\boldmath${\rm H}$}}
\newcommand{\veE}{\mbox{\boldmath${\rm E}$}}

\newcommand{\llan}{\langle\langle}
\newcommand{\rran}{\rangle\rangle}
\newcommand{\lan}{\langle}
\newcommand{\ran}{\rangle}
\newcommand{\vetau}{\mbox{\boldmath${\rm \tau}$}}

\begin{document}
\maketitle
\begin{abstract}
A new approach  to EW composite scalars  is developed, starting from the
fundamental gauge interaction on high scale. The latter is assumed to have the
group structure $SU(2)_L \times SU(2)_R\times SU(4)$ where $SU(4)$ is the
Pati-Salam color-lepton group. The  topological EW vacuum filled by instantons
is explicitly constructed  and the resulting equations for fermion masses
exhibit spontaneous $SU(2)$ flavor symmetry violation with possibility of very
large mass ratios.

\end{abstract}

\section{Introduction}

It is a rather common assumption, that the Electroweak Theory (EW) in its
standard version is an effective low-energy theory, produced by a fundamental
interaction at a high scale. If this fundamental theory exists, it should
answer many important questions, unanswered by the EW theory in its standard
form, for example:

\begin{enumerate}
    \item What is the dynamical origin of the Higgs field phenomenon and how
    the Electroweak Symmetry Breaking (EWSB),  proceeds in
    fundamental  terms from $SU(2)_W \times U(1)_Y$ to
    $SU(2)_W \times U(1)_{em}$, yielding masses of $W$ and $Z$.
    \item  What is the dynamical  mechanism producing  fermion masses.
    \item  What is the dynamical origin of generations.
    \item How one can explain the pattern of  fermion masses and mixings, in one generation  and
    in particular, a large difference between masses of quarks and neutrinos.
    \item  What is the origin   of the CP violation in EW.

\end{enumerate}
More questions can be added to that list,  e.g. why   at all the $SU(2)_L\times
U(1)$ structure appears, and then why it is broken by unequal masses of $t$ and
$b$ quarks and why left-right symmetry is broken. On the other hand,  many
physicists worked in the field last 50 years, producing a well-established and
an accurately checked picture of EW theory in good agreement with experiment
\cite{1}. Numerous efforts have been done to answer the first question,
suggesting dynamical models of composite Higgs mechanism. In a simpler form  it
was suggested to explain the composite Higgs field via the top condensation
mechanism (see \cite{2} for a review and references); in another version the
Technicolor model (TC) was suggested (see \cite{3} and \cite{4} for reviews and
references). In the extended versions (Extended Technicolor (ETC), Walking
Technicolor (WTC)) the Lagrangian contains both usual color, TC and flavor
interaction in one ETC gauge group \cite{4*}, for a recent review see
\cite{5*}.

The common to all these  approaches (actually trying to answer the points 1 and
2 above) is that a quark (plus possibly technoquark)  condensate is  formed at
a high scale, which gives mass to quarks and the same mechanism then provides
EWSB, yielding masses to $W$ and $Z$.

 As a consequence of TC and ETC models,  additional particles in the
 region of 1 TeV and higher was predicted, which can possibly be detected
 experimentally.

 To our knowledge no dynamical explanation for so different fermion masses and the structure of
 generations was given up to now in this type of  studies. Recently a new general  approach was suggested in \cite{5}, with a
 tentative  explanation of dynamical origin of generations  and the hierarchy of
 fermion masses. In particular, a simple pattern of masses  and  mixings,
 called the Coherent Mixing (CM) was developed for 3 and 4 generations and
 successfully compared to experimental data in \cite{6}.

 However, the group structure of the dynamical mechanism was not discussed in
 \cite{5,6}, neither the topic of EWSB was elaborated there, therefore we
 undertake below these two tasks and plan to  discuss in detail the change of
 group structure with the energy scale and the form of the corresponding
 effective Lagrangians. We assume below, that at some high scale only existing
 fermions are participating and new gauge interactions can appear, which are
 frozen at lower scales.

We start with one family --more families, as shown in \cite{5} can be produced
as additional solutions of the same dynamical equation.

Each family of fermions consists of 16 members $f_a^\alpha\equiv (u^\alpha_L,
d^\alpha_L, d_R^\alpha, u^\alpha_R)$, $\alpha=1,2,3$ (color), $\alpha=4 $--
leptons, so that $f^4_a= (\nu_L, e_L,  \nu_R, e_R)$. It is evident that indices
$\alpha, a$ can be organized into $U(4)_a \times U(4)_\alpha$ or $SU(4)_a\times
SU(4)_\alpha$ group indices. The $SU(4)_\alpha$ was introduced earlier in
\cite{7} as a unification of $SU(3)_{color}$ and lepton number. It is clear,
that $SU(4)_\alpha$ is splitted down to $SU(3)_{color}\equiv SU(3)_c$ at high
 scale, but we also know, that $SU(2)_W$ is symmertic to quarks and
leptons, and this calls for considering the possibility of $U(4)_\alpha$ or
$SU(4)_\alpha$ symmetry. We shall  have in mind the splitting pattern \be
 SU(4)_\alpha \to SU(3)_c \times U(1)_{B -L}\label{1}\ee

Concerning the $a$-indices, we shall consider the  flavor group $SU(4)_a \equiv
SU(4)_{EW}$. However, $SU(4)_a$ for a vector fundamental interaction
automatically breaks down to $SU(2)_L\times SU(2)_R$, and we shall study
explicitly the latter group, so that one essentially uses the group
$G(2,2,4)\equiv SU(2)_L\times SU(2)_R\times SU(4)$, introduced and exploited in
\cite{7},\cite{8*}.  The splitting pattern for that is similar to the standard
one \be E_6 ~{\rm or} ~ SO(10) \to G(2,2,4)  \to SU(2)_W \times U(1)_Y \to
SU(2)_W \times U(1)_{em}.\label{2}\ee In $G(2,2,4)$ each fermion
$\psi_{a\alpha}$ can have gauge interaction of three types: $A_\mu^{ab},
C_\mu^{a\alpha, b\beta}, B_\mu^{\alpha\beta}$.

The $B_\mu^{\alpha\beta}$. coincides with the usual color interaction when
$SU(4)_\alpha$ splits into  $SU(3)_c$. The first, $A_\mu^{ab}$ is the intrinsic
EW interaction at a  high scale producing, after the first splitting in
(\ref{2}), the gauge fields $W^A_\mu$ and $B_\mu$, and after EWSB, the fields
$W^A_\mu , A=1,2$ and electromagnetic $U(1)$ field. Now the field $C^{a\alpha,
b\beta}_\mu$ is local gauge field, which is adjoint both in $a,b$ and in
$\alpha,\beta$ indices, and the latter will be considered both in splitted and
unsplitted forms. We assume this interaction to be  active at high scale $M$ in
 an unbroken form, and surviving at low scale in the form of vacuum correlators, producing
nonzero average fermion bilinears and hence effective composite scalars, which
 give masses to fermions, and  in the course of EWSB also to vector fields.

We note  at this  point the possibility of inclusion into the game new
particles, like technofermions, which could interact also with the field
$C_\mu^{a\alpha, b\beta}$ and forming in this way the condensate of
technofermions, but this line will not be pursued further.

It  is clear, that the original $G(2,2,4)$ group is badly broken, and therefore
this should be reflected in the structure of vacuum averages of the fields
$\hat A_\mu, \hat C_\mu$.

In this way the properties of the vacuum fields and vacuum symmetries are
entering our problem together with the problem of the explicit mechanism of
symmetry violation. The related problem is that  of gauge invariance, since all
fields $\hat A_\mu, \hat B_\mu,\hat C_\mu$ are local gauge fields. As a general
statement, we shall assume always the mechanism of mass generation and symmetry
breaking to be associated with vacuum  field correlators, which can generate
scalar parts in the deconfined phase due to specific vacuum fields. An example
of such fields is given by instantons, which can produce masses and chiral
symmetry breaking (CSB), but do not give confinement at large distances. This
latter fact is due to exact mutual cancellation of all correlators of higher
powers in fields of this topological class at large distances \cite{8}, hence
no area law for the Wilson loop, but finite scalar contributions  in the
Green's functions of quarks at smaller distances, producing in this way masses
of quarks. Note, that this mechanism of CSB, studied before in \cite{11},  is
different from the one, generated by confinement of QCD \cite{9}.

As we shall see below, the quadratic correlator of the field
$C^{a\alpha,b\beta}_\mu$ can produce quartic combination  of  quark fields,
both of vector-vector and scalar-scalar kind. The latter  combination is
exactly what is  considered in topcondensate or TC models, however, as will be
seen, to  produce this combination of two {\bf white} bilinears (which is
necessary for fermion masses), one needs to start with the field
$C_\mu^{a\alpha, b\beta}$, depending on color indices $\alpha,\beta$ as an
adjoint  operator. Since bilinears depend also on $a,b$,  the same field $\hat
C_\mu $ should also contain these indices and be adjoint $SU(2)_{EW}$ field
before symmetry breaking. Another important outcome of the same mechanism,
composite vector fields appear on  the same ground as  the composite scalars,
and add to the possible intrinsic vector fields $\hat A_\mu$. Composite vector
fields have a long history \cite{*} (for a good discussion and references see
\cite{**}),  and have also been considered as candidates for EW gauge vectors
\cite{**}\cite{***}. This important topic needs a separate paper and will be
considered elsewhere.

The paper is organized as follows. In the next section the topological
structure of the EW vacuum is discussed and instanton ensembles are considered
as concrete examples.  The form of effective quartic quark Lagrangian is
derived in section 3  together with the effective Lagrangian of composite
vector gauge fields.  Spontaneous $SU(2)$ flavor symmetry breaking is discussed
in  section 4. In section 5  the  instanton-induced equation for fermion masses
is derived.
 In section 6 the obtained results are summarized.  Four appendices contain details of
derivations used in the paper.

\section{ Topological vacuum of electroweak theory}

As was discussed above, the central element of our construction (as well as in
topcondensate model \cite{2}) is the four-quark (and multiquark) effective
Lagrangian, which is produced by the field correlators of the fundamental gauge
fields. To have those field correlators one needs nontrivial nonperturbative
structure of the  vacuum, which  should establish the following properties of
resulting physical amplitudes:

\begin{enumerate}
    \item Local gauge invariance of the scalar self-energy of fermions, which
    ensures the physical fermion mass.
    \item Symmetry breaking of  the EW group  yielding finally
    $SU(2)_W\times U(1)_{em}$  with partially violated $SU(2)_W$, while
    $SU(3)_c$ and $B,L$ are not violated.
    \item Mass generation at low scale for all gauge fields involved,  except
    $SU(3)_c$, where  confinement is operating.

    \item CP violation.
\end{enumerate}

In the first item one needs a gauge invariant Chiral Symmetry Breaking (CSB)
phenomenon for an isolated fermion due to vacuum fields. It is known, that CSB
occurs due to scalar confinement and disappears in the QCD vacuum together with
it at $T\ga T_c$ \cite{9,10}. It is also known, that CSB may occur in the
instantonic model  of the QCD vacuum \cite{11},\cite{12},\cite{13}. As was
shown in \cite{9}, the difference  between these two cases of CSB can be
formulated in terms of the Field Correlator Method (FCM) \cite{14}, where the
fermion mass operator is represented as an  integral of a sum   over all
connected correlators $\ll F(1) ... F(n)\gg,$  $M=\sum_n M_n$,(see Appendix 1
for details). To establish gauge invariance, one starts with the gauge
invariant operator of the fermion mass in the  static field of heavy
antifermion at some fixed point $\veR_0$. It was shown in \cite{9}, that every
connected correlator ensures the linear term in the mass operator $M_n\sim c_n
|\vex -\veR_0 | + c_n^{(1)}$, where $c_n, c_n^{(1)}$ are some constants.

  It is crucial, what will be the result of summation
over $n$ \be M= |\vex -\veR_0| \sum c_n+ \sum c_n^{(1)}.\label{3}\ee
 In the QCD vacuum it is known, that the dominant contribution to the sum (\ref{3})
 comes from the lowest term with  $n=2$ \cite{8} and one obtains scalar
 confinement, which by itself implies CSB, and  confinement creates ``constituent mass'',
 which actually is the average energy of the confined quark \cite{15}. In  case
 of instantonic vacuum, all terms in the sum over $n$ in (\ref{3}) are
 important; moreover, as shown in \cite{8}, the sum  $\sum_n c_n$ vanishes
 for an ensemble of topcharges with integer fluxes, e.g. for instantons, and
 one obtains the finite value of $M^{(1)}=\sum c_n^{(1)}$, not depending on
 $\veR_0$, and hence fully gauge invariant. This is the case of CSB without
 confinement.

 Another  situation occurs  in QCD for $T\ga T_c$, where  the surviving nonconfining correlator $D_1$  produces the
 vector part of the
 quark selfenergy operator, and it appears in the form of the real part of the Polyakov
 loop \cite{16}  from the same gauge invariant construction  as discussed above, while confining correlators vanish at $T\geq T_c$.

 Thus the  requirement in the point 1 leads us to consider  the EW vacuum as
 ensemble of instantons (or more general solutions with integer fluxes) for the
 group $SU(2)\times SU(4)$, or more general subgroups of $E_6$ or $SO(10)$.

 Now the very structure of $SU(2)$ instanton solutions can help  to establish
 the symmetry breaking. Namely, in $4d$ gauge theory the basic element is the
 $SU(2)$ instanton of Belavin et al. \cite{17}, which can be embedded in the
 $SU(2)\times SU(4)$ construction in different ways.

 It is remarkable, that the resulting vacuum averages do not violate  $SU(2)\times
 SU(4)$, and the final phenomenon of spontaneous $SU(2)_{EW}$ violation will
 occur spontaneously  due to nonsymmetric fermion mass creation.

Finally, to establish CP violation, one must require, that the total density of
topcharges should be nonzero. For simplicity,  one can  assume, that vacuum
ensemble consists of only instantons (or antiinstantons), thus vacuum
condensate of gauge fields explicitly violates CP   (in addition this vacuum is
more stable, than instanton-antiinstanton vacuum). Then, using ABJ anomaly
relation, one can absorb the CP violating effect in the fermion phases of
(almost) massless fermions of the first generation.  In what follows, however,
we shall concentrate on the topic of fermion mass generation, leaving the
subject  of CP violation for another  publication.

We consider the $SU(2)$  instanton field in singular gauge with global color
orientation $\Omega$ \be A_\mu (x) =\bar \eta_{a\mu\nu} \frac{(x-R)_\nu
\rho^2\Omega^+ \tau_a \Omega}{(x-R)^2 [(x-R)^2 +\rho^2]}.\label{4}\ee Averaging
in the instanton  ensemble over each instanton is assumed with the weight
$D\gamma = \prod_{i=1}^N d\Omega_i \frac{dR_i}{V_4}$, and the following
equation for averaging in $d\Omega$ will be used \cite{18}  ($N_c =4$ for
$SU(4)$) \be \int d\Omega \Omega^+_{ab} \Omega_{cd} =\frac{1}{N_c}
\delta_{ad}\delta_{bc},\label{5}\ee and (here $a,\alpha$ refer to the same
group indices)
$$ \int d\Omega \Omega^+_{a\alpha } \Omega_{\beta b} \Omega^+_{a'\alpha'} \Omega_{\beta'b'} =\frac{1}{N_c^2-1} (\delta_{ab}\delta_{a'b'}
\delta_{\alpha\beta}\delta_{\alpha'\beta'}+ \delta_{ab'} \delta_{a'b}
\delta_{\alpha \beta'} \delta_{\alpha'\beta})-$$ \be-\frac{1}{N_c(N_c^2-1)}
(\delta_{ab}\delta_{a'b'} \delta_{\alpha\beta'}\delta_{\alpha'\beta}+
\delta_{ab'} \delta_{a'b} \delta_{\alpha \beta}
\delta_{\alpha'\beta'}).\label{6}\ee

Hence for the averaging of the square of instanton field of one can readily
deduce from (\ref{6}) \be \lan k^2\ran \equiv\int (\Omega^+ \tau^A \Omega)_{ab}
(\Omega^+ \tau^B\Omega)_{a'b'} d\Omega = \frac{tr(\tau^A \tau^B)}{N^2_c-1}
(\delta_{ab'}\delta_{a'b} -\frac{1}{N_c} \delta_{ab} \delta_{a'b'})\label{7}\ee

This can also be rewritten for $N_c=2$  as (since $tr (\tau^A\tau^B) = 2
\delta_{AB}$) \be \lan k^2\ran =  4/3t^C_{ab} t^C_{a'b'} \delta_{AB} ,~~
t^C=\frac12 \tau^C.\label{8}\ee

This can be immediately applied  to the $SU(2)_{EW}$ field $A^{ab}_\mu$, if one
consider the ensemble of $SU(2)$ instantons in the $SU(2)_{EW} $ vacuum,
$(A_\mu)_{ab}=\varphi^A_\mu (\Omega^+ \tau^A\Omega)_{ab}$ and averages over
color orientations the partition function (see  Appendix 2 for details of
derivation) \be Z=\int D\gamma D\psi D\bar \psi e^{i\int \bar \psi (\hat
\partial +m) \psi d^4 x+ \int \bar \psi \hat A \psi d^4 x}\label{9}\ee
 \be \lan e^{\int \bar \psi \hat A \psi d^4 x}\ran_\Omega =
e^{\sum^N_{i=1} \sum^\infty_{n=1} \frac{1}{n!} \ll\theta_i^n\gg},\label{10}\ee
where \be \theta_i \equiv \int \bar \psi \hat A \psi d^4 x \equiv \int \bar
\psi_{a\alpha} (\hat A)_{ab} \psi_{b\alpha} d^4x,\label{11}\ee and $\alpha$ are
$SU(4)_{lc}$ indices,  not participating in the averaging procedure. Applying
(\ref{7}), (\ref{8}) to $\ll \theta^2_i\gg$ and omitting the instanton index
$i=1,... N,$ on has

\be \ll \theta^2\gg = 2\int d^4xd^4y  \lan \varphi^A_\mu \varphi^A_\mu\ran_R
(\bar \psi_{a\alpha} (x) \gamma_\mu t^C_{ab} \psi_{b\alpha}(x)) (\bar
\psi_{a'\beta} (y) \gamma_{\mu'}t^C_{a'b'} \psi_{b'\beta}(y)).\label{12}\ee

One can see in (\ref{12}) the square of the effective vector field of the
$W_\mu$-type, which means, that $SU(2)$ instantons produce  effective vector
fields with the  unbroken symmetry. It is important, that color symmetry and
$SU(4)_{lc}$ are not broken, and  the effective $W_\mu$ field is color blind.
However at this stage the effective scalars do not appear and we must use
instantons in the field $C_\mu^{a\alpha,b\beta}$ to produce those.

We start with the averaging over fields $C^{a\alpha, b\beta}_\mu$ in the
quadratic effective Lagrangian, similar to $\ll\theta^2 \gg$ (\ref{12}), but
now in the ensemble of ``double instantons'' in $SU(2)\times SU(4)$ group,
which is proportional to

\be T\equiv \lan C^{a\alpha,, b\beta}_\mu C_\nu^{a'\alpha', b'\beta'}\ran_C
\sim \lan (\Omega^+ t^A \Omega )_{ab} (\Omega^+ t^B \Omega)_{a'b'}(\omega^+
\tau^D\omega)_{\alpha\beta} (\omega^+ \tau^E
\omega)_{\alpha'\beta'}\ran_{\Omega,\omega}\label{13}\ee

Using (\ref{7}), (\ref{8}), one obtains the following general structure
$$T=\delta_{\mu\nu}\left\{ \mathcal{M}_1\delta_{\alpha\beta
}\delta_{\alpha'\beta'} \delta_{ab}\delta_{a'b'}+\mathcal{M}_2
\delta_{\alpha\beta} \delta_{\alpha'\beta'}\delta_{ab'}\delta_{a'b}+\right.$$
 \be\left.
\mathcal{M}_3\delta_{\alpha\beta' }\delta_{\alpha'\beta}
\delta_{ab}\delta_{a'b'}+\mathcal{M}_4\delta_{\alpha\beta'}
\delta_{\alpha'\beta}\delta_{ab'}\delta_{a'b}\right\},\label{14}\ee which
should multiply the product of bilinears $S\equiv (\bar \psi_{a\alpha}
\gamma_\mu \psi_{b\beta}) (\bar \psi_{a'\alpha'} \gamma_\nu \psi_{b'\beta'})$.
Here $ \mathcal{M}_i = \mathcal{M}_i (x,y)$. For the product $TS$ one has to do
in the terms proportional to $ \mathcal{M}_3$, $  \mathcal{M}_4$ the Fierz
transformation to avoid colored bilinears,
$$TS= J_\mu J_\mu \left( \mathcal{M}_1+\frac12 \mathcal{M}_2\right) + 2 \mathcal{M}_2 J^A_\mu
J^A_\mu+ 2 \mathcal{M}_3 \sum_i c_i ( \bar \psi O_i t^A\psi) (\bar \psi
O_it^A\psi)+$$\be + \left(\frac12 \mathcal{M}_3+\mathcal{M}_4\right) \sum_i c_i
(\bar \psi O_i \psi) (\bar \psi O_i\psi). \label{15}\ee

Here  $J_\mu =\bar \psi \gamma_\mu \psi, ~~ J_{\mu A}= \bar \psi \gamma_\mu t^a
\psi$ and  $c_i=\left(-1,\frac12,\frac12,1 \right)$ for $S,V,A,P$ variants;
$O_i= (1,\gamma_\mu,\gamma_\mu\gamma_5,\gamma_5)$.

Note, that $a,b$ in (\ref{15}) run over a pair of $SU(2)$ indices and Dirac
$\frac{1\pm \gamma_5}{2}$ structure defines whether they belong to $SU(2)_L$ or
$SU(2)_R$. For the scalar current part Eq. (\ref{15}) one has $$ (TS)_{\rm
scal}= \sum_{i,k} \Lambda_{ik} \varphi_i \varphi_k^+= -4\left\{ (
\mathcal{M}_3+ \mathcal{M}_4)(\varphi_1 \varphi_1^++\varphi_2
\varphi_2^++\varphi_1\varphi_2^++\varphi_2\varphi^+_1)+
\right.$$\be\left.+\mathcal{M}_3(-\varphi_1 \varphi_2^+-\varphi_2
\varphi_1^++\varphi_3\varphi_3^++\varphi_4\varphi_4^+)\right\} \label{16}\ee

where we have defined $$ \varphi_1 =\bar u_R u_L, ~~ \varphi_2=\bar d_Rd_L,
~~\varphi_3 =\bar d_Ru_L, ~~ \varphi_4 =\bar u_R d_L,~~ \varphi_1^+= \bar
u_Lu_R,~~ {\rm etc. }$$

Finally, $\mathcal{M}(x,y)$ are  proportional to the trace of $(C^{a\alpha,
b\beta}_\mu(x) C^{a\alpha, b\beta}_\mu (y))$ averaged over positions and  sizes
of pseudoparticles, see Appendix 2  for details.

\section{Study of quartic quark operators}

As a result the partition function can be written as \be Z=\int D\psi D\bar
\psi e^{i\int \bar \psi \hat \partial \psi  dx + \frac12 \int\lan  (\bar \psi C
\psi)^2\ran_c{dxay}+...}\label{17}\ee In this form the quartic in $\psi,\bar
\psi$ and higher terms appear, which shall be treated below.

 As a next step, one  can use the simple
bozonization  procedure  to the vector terms in (\ref{16}), introducing
auxiliary vector field $v^{(i)}_\mu (x,y), i=1,2$. Using  the identities

 \be (\bar \psi J_i \psi) (\bar
\psi J_i\psi) = (v^{(i)} - (\bar \psi J_i \psi))^2 - (v^{(i)})^2 + 2
v^{(i)}(\bar \psi J_i\psi),\label{3.5} \ee $ i=\mu A, ~~ \mu A 5, \mu, \mu 5$,
$J_{\mu 5}\equiv \bar \psi\gamma_\mu \gamma_5 \psi,$ one can write the
partition function in the form
$$ Z=const \int D\psi D\bar \psi D v^{(i)}   \exp \left \{ i \int \bar \psi \hat
\partial \psi  d^4 x\right. +$$\be+ \left.\int dx dy \sum \xi_i (x,y)
[-(v^{(i)})^2+ 2 v^{(i)}(\bar \psi j_i \psi) ] + (ST)_{\rm scal} (\varphi,
\varphi^+)\right \}.\label{3.6}\ee Here $\xi_i$ are expressed via
$\mathcal{M}_i$ in  (\ref{14}), the explicit form will not be used  below.

 With respect to the last term in (\ref{3.6}) we apply the
procedure, used in \cite{5,6}, namely we introduce functional delta-function
and write \be e^{\int \Lambda_{ik }\varphi_i \varphi_k^+(x,y)  d^4x d^4y} =
\int\prod^4_{k=1} D\mu_k D\mu^+_k D\varphi_k D\varphi_k^+ e^{-K}\label{21}\ee
 with \be K= i \sum^4_{k=1}\int [\mu_k (\varphi_k-(\bar \psi \Gamma_k
\psi))+\mu_k^+(\varphi_k^+-(\bar \psi \bar \Gamma_k \psi))] dxdy + \int
\sum_{ik}\Lambda_{ik}\varphi_i \varphi_k^+ (x,y) dxdy\label{3.7}\ee and
$\varphi_k, \varphi_k^+$ and $\mu_k,\mu_k^+$ are to be found from the
stationary points of the total effective Lagrangian (see \cite{5} for more
details). As  a result due to (\ref{3.7}) quark fields now enter only in
bilinear forms in (\ref{21}), and one can integrate over them yielding the
final  effective Lagrangian (note, that  we also go over into the  Minkowskian
space-time (after all field averaging was done as always in the Euclidean
space-time), and we have put $\mu=\mu^+$ for simplicity, and  the dash sign to
show, that $\mu$ is a matrix $\mu_{ab}$. \be L_{eff} =\sum\xi_i (v_\mu^{(i)})^2
- \frac{1}{2} tr ln [(\hat \mu + i \hat D)(\hat\mu-i\hat D)],\label{3.8}\ee
where $D_\mu$ contains now vector auxiliary fields \be V_{1\mu}^{(A)}= 2 \xi_1
v_\mu^{(1)A}+2\xi_2 v_\mu^{(2) A}\gamma_5,~~V_{2\mu}= 2\xi_3 v_\mu^{(3)} +
2\xi_4 v_\mu^{(4)}\gamma_5,\label{3.9}\ee and can be written as \be iD_\mu
\equiv i\partial_\mu + V_{1\mu}^{(A)} t^A + V_{2\mu}.\label{3.10}\ee

To these terms in (\ref{3.9}) one could add intrinsic EW fields $g_2W^a_\mu t^A
+ g_1 B_\mu \frac{Y}{2},$ assuming that the  field $A^{ab}_\mu$ is broken to
this $(W,B)$ form by some other mechanism. We  can account for this
possibility, keeping in $V_{1\mu}, V_{2\mu}$ the corresponding intrinsic parts.

The operator under the logarithm in (\ref{3.8}) can be organized into several
forms (writing matrix $\hat m$ instead of $\hat\mu$ in case $\hat \mu\neq
\hat\mu^+)$ \be F\equiv (\hat m+i\hat D)(\hat m -i\hat D) =
\partial_\mu^2 + \hat m^2 + \Omega+ \hat N\equiv d^2 +\hat N + \hat
\Omega,\label{3.11}\ee where \be \hat N = \hat V\hat m -\hat m\hat V,~~ \hat V
\equiv (V_{1\mu}^A t^A+ V_{2\mu}) \gamma_\mu,~~\hat \Omega=-i(\hat \partial
\hat V + \hat V\hat
\partial) - \hat V^2.\label{3.12}\ee

Hence the  fermion loop expansion of $tr\ln F$ can be written as \be tr\ln F =
tr\ln [d^2(1+G(\hat N + \hat \Omega))]= tr\ln d^2 + tr (G(\hat N+\hat
\Omega))-\frac12 tr[G(\hat N+\hat \Omega) G(\hat
N+\hat\Omega)]+...\label{3.13}\ee where  matrix $G=(\hat m^2 +\partial^2)^{-1}$
and  after diagonalization of $\hat m\to \hat m_{\rm diag} (a)$, $G_a\equiv
(d^2)^{-1}_{aa} = (\hat m^2_a+\partial^2)^{-1}_{aa}$ is the free Green's
function of  fermion $a=1,2$ with mass $m_{\rm diag} (a)$, and higher terms of
expansion are neglected.

>From  (\ref{3.8}), (\ref{3.11}) it is clear, that $\hat m$ plays the role of
the fermion mass matrix, indeed for $\hat \mu\neq \hat \mu^+$ the mass operator
$\hat m$ in (\ref{3.11}) is \be\hat m \equiv \frac{\hat\mu+\hat\mu}{2} +
\frac{\hat\mu-\hat \mu^+}{2} \gamma_5, ~~ \hat \mu=\left(\begin{array}{ll}
\mu_1&\mu_4\\
\mu_3&\mu_2\end{array}\right)\label{28}\ee

$$\hat \mu^+=\left(\begin{array}{ll}
\mu_1^+&\mu_3^+\\
\mu_4^+&\mu_2^+\end{array}\right)$$

and $\mu_k,\mu^+_k$ can be obtained, as in \cite{5} differentiating
$\Lambda_{ik} \varphi_i\varphi_k^+$ in $\varphi_k,\varphi_k^+$ correspondingly,
which yields
\be \mu_i(x)= \int \Lambda_{ik} (x,y)  \varphi^+_k (y) dy; ~~ \mu^+_k (x) =\int
 \Lambda_{ik} (x,y)  \varphi_i (y) dy\label{29}\ee where the $\Lambda_{ik} (x,y) $
is easily obtained from (\ref{16}), \be \Lambda_{ik} =-4\left(
\begin{array}{cccc}
a&a-b&0&0\\a-b&a&0&0\\
0&0&-b&0\\
0&0&0&-b\\
\end{array}\right),\label{30}\ee
and  $a\equiv  \mathcal{M}_3+ \mathcal{M}_4$, $b=\mathcal{M}_3$.
 Both $\varphi_k, \varphi^+_k$
are to be found from $F$ (\ref{3.11}) as \be -i \varphi_k =\frac12
\frac{\delta}{\delta {\mu_k}} tr \ln F = \frac12 tr \left(
G\frac{\delta\hat m^2}{\delta\mu_k}\right), ~~ -i \varphi^+_k
=\frac12 tr \left( G\frac{\delta\hat
m^2}{\delta\mu_k^+}\right).\label{31}\ee Since both $G$ and $\hat
m^2$ are matrices, the connection of $\varphi_k$ and $\mu_l$ takes
the form  in the momentum space (see Appendix 3 for  a derivation)
\be - i \varphi \equiv  d(p) = \frac{\mu^+(p)}{2(p^2+\mu\mu^+)},~~
-i\varphi^+ =d^+ (p) = \frac{\mu(p)}{2(p^2+\mu\mu^+)}
\label{32}\ee therefore  the final form of equations for $\hat
\mu, \hat \mu^+$ is \be \mu_i(p) = \int K_{ik} (p,p_1) \mu_k(p_1)
d^4p_1 + ... \label{33}\ee $$\mu^+_{i}(p) = \int\bar K_{ik}
(p,p_1) \mu^+_k(p_1) d^4p_1+...$$ and e.g. $K\sim \Lambda_{ik}$
the dots signify contribution of higher powers of $\mu, \mu^+$.
Thus one obtains in general nondiagonal $SU(2)$ matrices for $\hat
\mu, \hat\mu^+$ and $\hat V $; moreover, since $\hat V$ and $\hat
m$ contain $\gamma_5$, the interaction is different for left and
right particles, so that the resulting form of (\ref{33} )
violates both $SU(2)$ flavor and left-right symmetry, as will be
discussed in the next section.

\section{Electroweak symmetry breaking}

There are two facts of EW symmetry breaking,

1) $SU(2)_{EW} \times U(1)_Y\to SU(2)_{EW} \times U(1)_{em}$, and 2) breaking
of $SU(2)_{EW}$ by unequal mass terms of up and down fermions. We shall
demonstrate below, that both types of EWSB are present in the resulting
$L_{eff}$ (\ref{3.8}). We start with the point 2), and remark, that the form
(\ref{32}) with nondiagonal matrices  $\hat K, \hat K^+$ implies, that
eigenvalues $\hat \mu,\hat\mu^+$ are not equal. Moreover, one may have
asymmetric in $\mu,\mu^+$ solutions, which automatically leads to the CP
violation. To study fermion mass eigenvalues of the matrix $\hat m$, we first
simplify to the case $\hat\mu=\hat\mu^+$, where $\hat m$ has the form
(\ref{28})  in the $(\bar u, \bar d)\times (u,d)$ basis, so that the
eigenvalues of  $\hat m=\hat \mu$ are easily found to be \be \bar m_{1,2} =
\frac{\mu_1+\mu_2}{2} \pm \sqrt{\left(
\frac{\mu_1-\mu_2}{2}\right)^2-\mu_3\mu_4}.\label{34}\ee

Now $\mu_i$ are to be defined  from (\ref{33}), and one finds matrices $K,\bar
K $ etc. from (\ref{31}). In the case $\hat \mu=\hat\mu^+$ one derives a system
of equations (see Appendix 3 for details of derivation)$\mu_i= \int
\Lambda_{ik} d (\mu_k)$ and we omit the integration signs,
$$\mu_1=-4[(\mathcal{M}_3+\mathcal{M}_4) d(\mu_1) + \mathcal{M}_4 d(\mu_2)]$$
$$\mu_2=-4[(\mathcal{M}_3+\mathcal{M}_4) d(\mu_2) + \mathcal{M}_4 d(\mu_1)]$$
\be \mu_3=-4\mathcal{M}_3 d(\mu_4)\label{35}\ee
$$\mu_4=-4\mathcal{M}_3 d(\mu_3),$$ where $d(\mu_i)=
\frac{\mu_i}{p^2+m^2}$.

Hence the symmetry of the matrix $\Lambda$ defines the symmetry of eigenvalues
$\mu_i$. In particular, when $\mu_3$ and /or $\mu_4$ vanishes, one
automatically obtains $\mu_1=\mu_2=\bar m_{1,2}$ and no $SU(2)$ flavor
violation results. In the specific case, when $a=0$ in (\ref{30}), one has
$\mu_1 =\mu_2=\mu_3=-\mu_4$, and  $\bar m_1=2\mu_1, \bar m_2=0$, exemplifying
the maximal flavor violation. It is important to note, that by diagonalization
of $\hat m$ one defines the ``true'' up-and down fermions, which are rotated
with respect to original $u,d$ states.

 Now we turn to the point 1).

It is useful to represent vector potential $V_\mu$ (\ref{3.10}) in the $(L,R)
\times $ (up, ~ down) matrix notations, so that  $$V_\mu =(V_L^A + V_R^A )t^A +
V_L\hat 1 + V_R\hat 1 \equiv \bar V_L + \bar V_R=$$\be=\left(\begin{array}{ll}
V_L^At^A+V_L\hat 1,& 0\\0,&0\end{array} \right)+ \left(\begin{array}{ll}
0,&0\\0,&V_R^At^A+V_R\hat 1\end{array} \right)\label{34}\ee and $\hat m =
\left(
\begin{array}{ll} o&\hat \mu^+\\\hat\mu&0\end{array}\right)$,  so that (for
$\hat \mu=\hat \mu^+)$ one obtains \be \hat N= \hat V \hat\mu-\hat\mu \hat
V=\left( \begin{array}{ll} 0, & \bar V_L\hat \mu-\hat \mu\bar V_R\\\bar V_R\hat
\mu-\hat \mu\bar V_L,&0\end{array}\right).\label{35a}\ee

The basic role in the EWSB is played by the last term in (\ref{3.13}), which
can be rewritten as (here $tr \equiv \frac14 tr_D tr_{a,b}$, subscript $D$
refers to Dirac matrices $\gamma_\mu$).\be tr\ln F = ... -\frac12 [tr (G\hat N
G\hat N)+tr (G\hat \Omega G\hat \Omega)].\label{3.14}\ee As a first example we
take for illustration the case, when $\hat V= V_L^At^A + V_2 y(a) $, one
obtains  in the  standard way \cite{2,24} \be tr (G\hat N G\hat N)
=-\frac12(V^A_{1\mu})^2 (\mu^2_u+\mu^2_d) G_uG_d-\frac12 [2(y_1-y_4) V_{2\mu} +
V^3_{1\mu}]^2 (\mu^2_u G^2_u+\mu^2_d G^2_d)\label{3.15}\ee where we have
introduced the hypercharge $Y_a$ of the fermion $a$, \be y_{ab} =y(a) = diag
(y_1, y_2, y_3, y_4),~~ y_a =\frac{Y_a}{2} \label{3.16}\ee

Here $G^2$ is actually the integral cut-off at large scale $M$ \be G^2_{u,d}
\to \frac{1}{(2\pi)^4} \int
\frac{d^4p}{(p^2+\mu^2_{u,d})^2}=\frac{1}{16\pi^2}\int^{M^2}_0
\frac{p^2dp^2}{(p^2+\mu^2_{u,d})^2}.\label{3.18}\ee

 A more general case, EWSB, when both $\hat V_L$ and $\hat V_R$ are present in
 $\hat N $ (\ref{35a}), is considered in Appendix 4. In this case additional
 terms    appear in the resulting effective action making the whole picture of
 EWSB more complicated, which  will be treated elsewhere \cite{22}.

\section{Properties of selfconsistent solutions}

In this section we analyze  the properties of solutions  for mass eigenvalues
of Eq. (\ref{35}), using   Eq.(\ref{A16}) for $\mu (p)$, tfor he kernels $\bar
J_n$, obtained in Appendix 2 for the randomized ensemble of instantons. We also
distinguish $\mu$ and $\mu^+$, since (as also shown in Appendix 2) topological
zero modes have definite chirality and produce different contribution to $\mu$
and $\mu^+$. The resulting equations can be rewritten from (\ref{A16}) as (all
$\mu, \mu^+$, $d, d^+$ are matrices in flavor space).
$$ \mu(p) = \int \frac{d^4p_1}{(2\pi)^4}\bar J_2 (p,p_1) d^+ (p_1) -\int
\prod_{i=1,2,3} \frac{d^4p_i}{(2\pi)^4} \bar J_4 (p,p_1, p_2, p_3)\times$$ \be
\times   d^+ (p_1) d(p_2) d^+(p_3) + \int \prod^5_{i=1} \bar J_6 (p,p_1,...p_5)
d^+ (p_1) d(p_2)... d^+(p_5)+...\label{43a}\ee
 For $\mu^+(p)$ one should replace on the r.h.s. $d^+\leftrightarrow d$. Here
 $\bar J_n$ in the randomized instanton ensemble have the form (see Appendix 2
 for details)
 \be \bar J_2 (p,p_1) =\frac{N_{top} k^2(q)q^2}{V_4 N_c^2 (2\pi)^4} = \left(
 \frac{\rho}{R}\right)^4
 \frac{\chi^2(p-p_1)}{N^2_c(2\pi)^4(p-p_1)^2},\label{44}\ee
 where $$\chi(q) =
 \left\{ \begin{array}{ll} -\frac12,& q\to 0\\ -2/(q\rho)^2,&
 q\to\infty,\end{array}\right. {\rm ~ and}~ k(q) =\frac{\chi(q)}{q^2}$$
 \be \bar J_4 (p,p_1,p_2,p_3) = \frac{N_{top}}{4V_4} \left(
 \frac{4\rho^4}{(2\pi)^4}\right)^2 \prod^4_{i=1} k(q_i)(q_1q_2)
 (q_3q'_4),\label{45}\ee
 and  $q_1=p-p_1, q_2=p_1-p_2, q_3=p_2-p_3, q'_4=p -p_3, q_4=-q'_4,$.

 The Eq.(\ref{43})  allows to find solutions for $\mu(p)$ and $\mu^+(p)$, since
 $d(p)$ and $d^+(p)$ are expressed via $\mu, \mu^+$: in the simple
 approximation $\mu=\mu^+$ one has $d(p)=d^+(p)= \frac{\mu(p)}{p^2+\mu^2(p)}$
 and in case of  right zero mode contribution, $d(p)$ and $d^+(p)$ are given in
 (\ref{A30}) and (\ref{A31}) respectively.

As a next point, in this section we discuss the general structure of the
spectrum of Eq. (\ref{43a}) using the random  instanton vacuum (RIV) as an
example. In this case, taking $J_n(p, p_1,...p_{n-1})$ from  Appendix 2,
Eqs.(\ref{A14}), (\ref{A16}) one can rewrite Eq. (\ref{43a}) via dimensionless
combinations

\be\mu\rho=\frac{\rho^4}{4R^4}\int\prod^{n-1}_{i=1} \left( \frac{d^4p_i
(\mu(p_i)\rho)}{q^2_{i+1}(p^2_i +\mu^2(p_i))}\right) \prod_{k=1,3...}^{n-1}
\left[ \frac{4\rho^2 (q_k q_{k+1})
\chi_k\chi_{k+1}}{\sqrt{N_c}}\right]\label{56} \ee

Here $\rho$ is the instanton size parameter, $R^4=V_4/N_{top}$ is the inverse
instanton density, $q_i = p_i -p_{i+1}, ~~p_{n+1}\equiv p_1$. One can see, that
$\chi_k\cong -\frac{2}{(q_k\rho)^2}, ~~ |q_k\rho|\to \infty$, plays the role of
the cut-off factor, and in the  region of large $p_i$ the integrand is well
behaved, while in the infrared region $(p_i\to 0)$ the first term on the r.h.s.
is logarithmically divergent for $\mu\to 0$, while all others are IR safe.
Moreover, it is clear, that the n-th term is proportional to $(\mu\rho)^{n-1}$,
and since $\rho\sim 1/M$, one expects, that $\frac{\mu}{M}\ll 1$ for all roots
of (\ref{56}). Hence all terms with $n>2$ in this case of RIV should be
subleading and can be omitted in the first approximation.

We shall concentrate now on the first term $J_2$ in (\ref{43a}) and rewrite it
for the RIV as (assuming at this stage $\mu(p) = \mu^+(p)$ and omitting  flavor
matrix indices) \be \mu(p) = b\int \frac{\mu(p_1) d^4p_1}{(p-p_1)^2
(p^2_1+\mu^2(p_1))(2\pi)^4}\label{57}\ee with $b=\frac{\rho^4}{4N_cR^4}$, and
having in mind the cut-off at large $p_1 \approx 1/\rho$ due to the suppression
factor $\chi^2(p-p_1)$, not shown in (\ref{57}). As a first guess one can put
$\mu(p) =\mu_0 =const\ll 1/\rho$, and obtains $(M\equiv 1/\rho)$ \be 1\cong
\frac{b}{16\pi^2} ln \frac{M^2}{\mu^2_0}, ~~ \mu^2_0 = M^2 \exp \left(
-\frac{16\pi^2}{b}\right).\label{58}\ee

One can see, that for small $\frac{b}{16\pi^2}\ll 1$ the resulting values of
$\mu_0$ can be very small, so that (\ref{58}) illustrates the mechanism of
small fermion mass due to RIV. Note, that the instanton solution was important
in this, since it produces additional factor $\frac{1}{(p-p_1)^2}$ in
(\ref{57}).

In appendix 2, one can  see  that the same type of factor appears for the case
$\mu\neq \mu^+$ due to zero mode contributions. Note, that the logarithmic
instability of (\ref{57}) is a typical feature, which is revealed in the $3d$
nonrelativistic potential $1/r^2$.  One  expects in this case, that there are
additional (excited) states, provided the strength of potential is large
enough.  It is important to study possible mechanisms of relaxation at large
distances in RIV,  which exist for the  mixture of instanton  and
antiinstantons. However, for the homogeneous instanton gas without
antiinstantons one can have a sum of individual contributions without
interaction between instantons, and thus the relaxation  may be absent.

Now  we restore the  flavor indices and consider Eq. (\ref{58}) diagonalized in
the flavor space, so that $\mu(p) \to \mu_i(p), b\to b_i, i=1,2$. As can be
seen in (\ref{57}), the small and uniqual $b_1\neq b_2$ can lead to  very
different $\mu_0^{(i)}, \mu_0^{(i)}= \mathcal{M}\exp \left(
-\frac{8\pi^2}{b_i}\right)$ thus strongly amplifying the disparity of
coefficients $b_i$ defined by $\Lambda_{ik}$ (the ``zoom effect'').

\section{Summary and conclusions}

We have obtained above two basic results. First, we have derived equations for
fermion masses $\mu_k, \mu_k^+$, Eq. (\ref{33}),  which defined
selfconsistently both diagonalized (physical) fermion masses $\mu_d$, $\hat \mu
= U\mu_d U^+,~~ U=a+i\ven\vetau$, and $\mu^+_d$. The equations contain the
kernel $\mathcal{M} (x,y)$, which is proportional to the quadratic field
average  $\lan C_\mu(x)C_\mu(y)\ran$, and  can be explicitly calculated for the
random instanton ensemble.  As it is, the nondiagonal structure of kernels
$K_{ik}$ etc., already seen in (\ref{30}), presupposes $SU(2)$ violation in
fermion masses, revealed in unequal masses of up and down fermions. Note, that
our original interaction $C_\mu^{a\alpha, b\beta}$ and its quadratic average do
not violate $SU(2)_L$ or $SU(2)_R$, and this violation occurs spontaneously in
the final equations  (\ref{35}) resulting from the stationary points of the
effective action (\ref{12}). Another basic  result  of the  present paper is
the  integral equation  for fermion mass eigenvalues (\ref{57}), where the
kernel is provided by instanton density and shape, yielding approximate
solutions of the form (\ref{58}), highly sensitive to the  small differences in
the kernel values for two diagonal mass eigenvalues.

This latter equation (\ref{58}) is able in principle provide large ratios of
masses in the same $SU(2)$ multiplet, as in $m_t/m_b$, and in ratios of quark
to lepton masses. This type of analysis is planned for the future. A different
scenario to explain large mass ratios and $SU(2)$ flavor violation was
suggested recently in \cite{26}, where a new type of horizontal interaction was
introduced.

The results of the present study can be considered also in the
framework of the left-right symmetric models  \cite{27,28}, with
possible introduction of the effective composite scalars instead
of elementary Higgs field. Another useful connection could be with
the recently proposed model \cite{29}, where left-right symmetric
flavor symmetry is broken by the Yukawas. Finally, effective
composite scalar of the kind considered in the present paper could
be exploited in the GUT scenarios, based on $SO(10)$ gauge
symmetry  \cite{30,31},  where important quantum effects \cite{32}
can be formulated in terms of composite fields.

The author is grateful to A.I.Veselov for discussions, the financial support of
RFBR grant no. 09-02-00620a is gratefully acknowledged.



\vspace{2cm}
 \setcounter{equation}{0}
\renewcommand{\theequation}{A1.\arabic{equation}}

{\bf \large
\noindent Appendix  1}\\

\noindent{\bf \large Chiral symmetry breaking and fermion mass generation }\\

We start with the case of heavy quarks and consider the Green's function of a
white pair of static quark  and antiquark $Q\bar Q$ at distance $R$, which is
proportional to the Wilson loop $(R\times T$), which can be written as a sum
over the field correlators (cluster expansion) \cite{8,14} $$ W(S) =\lan \exp
(ig \int_S ds_{\mu\nu} F_{\mu\nu} (z))\ran= $$\be =\exp \sum_n
\frac{(ig)^n}{n!} \int_S ...\int_S ds_{\mu_1\nu_1} (u_1)
ds_{\mu_2\nu_2}(u_2)... d s_{\mu_n \nu_n} (u_n)\times\label{A2.1a}\ee $$\times
D_{\mu_1\nu_1...\mu_n\nu_n} (u_1 , u_2,...u_n).$$ We shall be interested only
in the confining correlators, which can be written as

\be
  \llan F_{\nu_1\mu_1}(u^{(i)})...
   F_{\nu_n\mu_n}(u^{(n)})\rran =
   \prod_{i,k}(\delta_{\nu_i\nu_k}\delta_{\mu_i\mu_k}-
   \delta_{\nu_i\mu_k}\delta_{\nu_k\mu_i})D^{(n)}(u^{(i)},...
   u^{(n)})\label{A2.4}
   \ee

Separating the c.m. coordinates, yielding total area $||S||$ of
the surface $S$, one has for the flat minimal surface \be W(S)
=\exp (-||S||\sum - (R+T)\kappa+ O(R/T, T/R)),\label{A2.3a}\ee
$$ \sum=\sum_{n=2k} \frac{(ig)^n}{n!} \int d^2 u_1... d^2u_n
D^{(n)}(u_1,...u_n)$$ where $D^{(u)}$ depends only on relative coordinates
$(u_i-u_j)$, which change in the intervals $(-\infty,\infty)$. Now for the
field correlators of the vacuum configurations consisting of instantons with
integer fluxes the sum $\sum =0$ in (\ref{A2.3a}), implying no confinement.
However, there exist perimeter terms, (do not confuse these nonperturbative
terms with diverging perturbative artefacts, specific for the Wilson loop of
static  quarks). These terms arise from the finite intervals of integration in
$\sum$ in (\ref{A2.3a}). E.g. for the quadratic correlator $D^{(2)}$ one has
the static potential \be V_{Q\bar Q} (R) = 2 \int^R_0 (R-\lambda) d\lambda
\int^\infty_0 d\nu D^{(2)} (\sqrt{\lambda^2 +\nu^2})\label{A2.4a}\ee and   for
large $R$ one obtains \be V_{Q\bar Q} (R\to \infty) =\sigma R +
d_2,\label{A2.5a}\ee with $\sigma = 2 \int^\infty_0 d\lambda \int^\infty_0 d\nu
D^{(2)} (\sqrt{\lambda^2 +v^2}), ~~ d_2 = - 2\int^\infty_0 \lambda d\lambda
\int^\infty_0 d\nu D^{(2)} (\sqrt{\lambda^2 +v^2}).$

The sum over $n$ of constant terms need not vanish, and one can write for the
ensemble of integer fluxes \be V_{Q\bar Q} (R\to \infty) =-\sum^\infty_{k=0}
\frac{(-g^2)^k}{(2k)!} \int^\infty_0 \lambda_1 d\lambda_1... \lambda_{2k-1}
d\nu_1... d\nu_{2k-1}\times \label{A2.6a}\ee $$ D^{(2k)} (\lambda_1, \nu_1,...
\lambda_{2k-1}, \nu_{2k-1})$$ thus one obtains a finite scalar contribution to
$V_{Q\bar Q}(R)$ which  does not depend on $R$ at large $R$ and one half of it
can play the role of the effective mass of each of the static quarks, (which
can be also negative). However, it was derived for static quarks, and we need
equivalent expressions for fermions of any mass. To do that we shall exploit
another technic, useful for the system of light fermion of any mass in the
field of static antifermion.

Below we give the gauge invariant derivation of the fermion
self-energy part in terms of field correlators, following the
reference \cite{9}.

We start with the calculation of effective Lagrangian for the
system of a colored fermion in the field of a static antifermion
at point $\veR_0$. It can be written in the leading order of the
$1/N_c$ expansion as follows: \be \lan \exp\left[  \int \bar
\Psi_\alpha \hat B^{\alpha\beta} \Psi_\beta d^4 z \right]\bar \psi
(x) P\exp (ig \int^x_y B_\mu (u) du_\mu) \psi (y)\ran_{B,\psi,
\bar \psi}\label{A2.1}\ee

The averaging over $B_\mu$ and $\psi,\bar\psi$ yields the cumulant expansion in
the exponent  and brings about the following form of the self-energy part in
the fermion-antifermion Green's function,
$$
iM^{(n)}(x^{(1)},...x^{(n)})=\gamma_{\mu_1}S(x^{(1)},x^{(2)})
\gamma_{\mu_2}...\gamma_{\mu_{n-1}} S(x^{(n-1)},x^{(n)})\gamma_{\mu_n}\times
$$
\be \times N^{(n)}_{\mu_1... \mu_n}(x^{(1)},...x^{(n)}) \label{A2.2} \ee where
we have defined
$$
 N^{(n)}_{\mu_1... \mu_n}= \int^{x_1}_0 d\xi^{(1)}_{\nu_1}
  \int^{x_2}_0 d\xi^{(2)}_{\nu_2}...
  \int^{x_n}_0 d\xi^{(n)}_{\nu_n}
  \alpha
  (\xi_{\nu_1})...
  \alpha
  (\xi_{\nu_n})\times
  $$
  \be
  \ll F_{\nu_1\mu_1}(\xi^{(1)})...
   F_{\nu_n\mu_n}(\xi^{(n)})\gg\label{A2.3}
   \ee
   and $  \alpha(\xi_4)=1,          ~~
    \alpha(\xi_i^{(k)})=
    \frac{\xi_i^{(k)}}{x^{(k)}_i},~~
    i=1,2,3; k=1,...n.$ Here $\vexi^{(i)}=0$ at $\veR_0$.

    One  can identify in cumulant $\ll...\gg$ the part similar to
    $D$, i.e. violating the Abelian Bianchi identity, namely
    for even  $n$
  \be
  \llan F_{\nu_1\mu_1}(\xi^{(1)})...
   F_{\nu_n\mu_n}(\xi^{(n)})\rran =
   \prod_{i,k}(\delta_{\nu_i\nu_k}\delta_{\mu_i\mu_k}-
   \delta_{\nu_i\mu_k}\delta_{\nu_k\mu_i})D^{(n)}(\xi^{(1)},...
   \xi^{(n)})\label{A2.4}
   \ee
Note, that the total Green's function of the fermion-static antifermion system
is gauge invariant and for in and out coordinates $x,y$ in (\ref{A2.1}) can be
written as \be G(x,y,\veR_0) =tr (\Phi(x,y,\veR_0) S(x,y))\label{A2.5}\ee where
\be \Phi(x,y,\veR_0) =\Phi(\vex, x_4; \veR_0, x_4) \Phi(\veR_0, x_4; \veR_0,
y_4)\Phi(\veR_0, y_4; \vey, y_4)\label{A2.6}\ee and $\Phi$ is the parallel
transporter \be \Phi(a,b) = P\exp (ig \int^a_b B_\mu du_\mu)\label{A2.7}\ee

Let us consider now the case of light quark $q$ in the field of the static
antiquark $\bar Q$. The initial and final states  can be written in the
gauge-invariant form as \be \Psi_{in,out} (x,y) = \bar \psi_q (x) \Phi(x,y)
\Psi_{\bar Q} (y).\label{A2.8}\ee

The Green's function for the total $q\bar Q$ system can be written as
$$ G_{q\bar Q} (x,y|x',y') = \lan \Psi^+_{out} (x', y') \Psi_{in}
(x,y)\ran_{q,B} =$$ \be = \lan tr (\Phi(x',y') S_Q (y',y) \Phi(y, x) S_q(x,
x')\ran_B,\label{A2.9}\ee and for the static quark one can take $S'_{q} (y', y)
\sim \Phi(y', y)$ where $\Phi(y',y)$ is along the straight line, and for $S_q$
one can use the  Fock-Feynman-Schwinger Representation (FFSR) {}, \be S_q (x,y)
= (m-\hat D) \int^\infty_0 ds (Dz)_{xy} e^{-K} P_B \exp \left( ig \int^x_y
B_\mu dz_\mu\right) p_\sigma (x, y; s)\label{A2.10} \ee Here $\hat D =\hat
\partial - ig \hat B,$ $P_B$ is the ordering operator, and  $p_{\sigma}$ is
spin-dependent factor.

\be p_\sigma (x,y;s)= P_F \exp [ g \int^s_0 d\tau \sigma_{\mu\nu} F_{\mu\nu}
(z(\tau))]\label{A.11}\ee and $\sigma_{\mu\nu} F_{\mu\nu}$ is \be
\sigma_{\mu\nu} F_{\mu\nu} =\left( \begin{array}{ll}\vesig \veH,& \vesig \veE\\
\vesig \veE,& \vesig\veH\end{array} \right)\label{A.12}\ee As a result $G_{qQ}$
can be written in a simpler form \be G_{q\bar Q} (x,y| x',y')=\int^\infty_0 ds
(Dz)_{xx'} e^{-K} \lan tr (m-\hat D) W_\sigma (x,y| x',
y')\ran_B\label{A.13}\ee and $W_\sigma (x,y| x',y')$ is the Wilson loop with
spin insertions of $(\sigma F)$ factors and with contour $C(x,y|x'y')$
consisting of the variable path from $x$ to $x'$ and three straight-line pieces
$\Phi(x',y')\Phi(y',y) \Phi(y,x)\equiv \Phi(x', y', y, x)$ \be W_\sigma (x,y|
x',y')= tr [\exp (ig \int^x_{x'} B_\mu dz_\mu) p_\sigma (x,y;s) \Phi (x', y')
\Phi(y', y) \Phi(y, x)]\label{A.14}\ee

Writing (\ref{A.13}) as a gauge invariant combination, \be G_{q\bar Q} =\lan tr
\Phi (x', y', y,x) S_q (x,x')\ran_B\label{A2.15}\ee one can see, that we are
interested in the situation, where there is Chiral Symmetry Breaking (CSB) so
that $\lan S_q(x,x)\ran$ is nonzero, but  confinement is missing, so that one
can remove the heavy antiquark to infinity. To get this property, one starts
with the ensemble of instantons, and keeping in the cluster expansion of the
Wilson loop $W_\gamma$ only the quadratic (Gaussian) term, one obtains
confinement and CSB at the same time (see  \cite{9} for details). However,
taking all correlators, one gets as in previous example of Wilson loop  for
static quarks, no linear term in the $q\bar Q$ interaction -- no confinement,
but may have remnants of scalar  interaction, as in \ref{A2.4a}. This is enough
for CSB, since appearance  of scalar pieces in the effective interaction
(effective mass) signals CSB. Thus one obtains CSB for the ensemble of
instantons
 in absence of confinement. This phenomenon of CSB in instantonic vacuum was
studied before in \cite{11}.

\vspace{2cm}
 \setcounter{equation}{0}
\renewcommand{\theequation}{A2.\arabic{equation}}

{\bf \large
\noindent Appendix  2}\\

\noindent{\bf \large Field correlators of a random instanton ensemble}\\

 We consider here the vacuum of  SU(2) gauge theory filled by the
 noninteracting gas of instantons  I (and possibly  antiinstantons
\={I}), so that the total vector potential $A_\mu$ is

\be A_\mu(x) = \sum^N_{i=1}  A_\mu^{(i)} (x, \gamma_i)\label{A1}\ee where
$\gamma_i$ defines the set of collective  coordinates: position, color
orientation and size $ \gamma_i = \{ R^{(i)}, \Omega_i, \rho^{(i)}\}$, and we
are interested in the effective fermion Lagrangian, which obtains after
integration of partition function over all collective coordinates
 \be Z=\int d\gamma D\psi D\bar \psi e^{\int \bar \psi (i \hat
 \partial + im + g\hat A) \psi dx}= \int d\psi d\bar \psi e^{\int \bar \psi (i \hat
 \partial + im ) \psi  + L_{eff} (\psi, \bar \psi)}.\label{A2}\ee

 Here notation is used
 \be d\gamma = \prod^N_{i=1} \left( d\Omega_i \frac{d^4
 R^{(i)}}{V_4}\right)\label{A3}\ee
 and the instanton  potential in the singular  gauge is
 \be
 A_\mu^{(i)} (x,\gamma) = \bar \eta_{a\mu\nu}\frac{(x-R^{(i)})_\nu
 \rho^2 \Omega^+_i \tau_a\Omega_i}{(x-R^{(i)})^2 [ ( x- R^{(i)})^2
 +\rho^2]}.\label{A4}\ee

 It was shown in \cite{13} that $L_{eff}(\psi, \bar \psi)$ can be
 found by the  cluster expansion in the limit of large $N_c$ (when
 SU(2) is the subgroup of $SU(N_c))$ and when $N, V_4\to \infty,
 ~~ \frac{N}{V_4}=const$ (the thermodynamical limit).

 The result for $L_{eff}$ is the sum over instantons $(i)$ and
 over power $n$ of cumulant  $\ll (A_\mu)^n\gg$, \cite{13},
 \be L_{eff} =\sum^N_{i=1} \sum^\infty_{n=2} \frac{1}{n!} \ll
 \theta^n_i\gg\equiv \sum^N_{i=1} L_{eff}^{(i)}\label{A5}\ee
 where we have denoted ($f_i$--flavor indices).
 \be
 \theta_i =\sum^{N_f}_{f_i=1} \theta_i^{f_i},
 ~~\theta_i^{f_i}\equiv\int dx \psi^+_{f_i}(x) \hat S^{(i)}
 \psi_{f_i} (x).\label{A6}\ee

 The calculation in \cite{13} yields the following answer
 \be L_{eff}^{(i)} =\sum^\infty_{n=2} \frac{ N}{2nV_4 N_c^2}
 \prod^n_{k=1} \int O_k \frac{dp_k dq_k}{ (2\pi)^8} (2\pi)^4
 \delta (\sum^n_1 q_k) tr \{ \prod^n_{j=1} A_{\mu_j}
 (q_j)\}\label{A7}\ee
 where
 \be O_k \equiv \psi^+_{\alpha_k} (p_k) \gamma_{\mu_k}
 \psi_{\beta_k} (p_k-q_k) \delta_{\alpha_k,
 \beta_{k-1}}.\label{A8}\ee

 For our purpose we need to form white bilinears $(\psi^+_\alpha
 \psi_\alpha)$ using pairwize Fierz identities, keeping only
 scalar and pseudoscalar combinations  and taking into account
 antisymmetry of $\psi, \psi^+$.

 This yields
\be \prod ^n_{k=1} O_k \to - \prod^n_{k=1,3,...} \Phi^{(t_k)}_{LR}
 (p_{k+1}, p_k-q_k) \Phi_{RL}^{(t_k)} (p_{k+2}, p_{k+1}
 -q_{k+1})\label{A9}\ee
 where $\Phi_{LR}$ denotes a  fermion-antifermion pair,
 \be \Phi_{LR} (p', p-q) = \psi^+_{Lf'}(p')\psi_{Rf} (p-q), ~~
 t=(f'f).\label{A10}\ee

 For the vacuum-averaged pair $\lan \Phi_{LR}\ran $ one has the
 property
 \be \Phi^{(t)}_{LR} (p', p-q)= \delta (p'-(p-q))
 \varphi^{(t)}_{LR} (p') .\label{A11}\ee

 As a result one can write $L_{eff}^{(i)}$ in the form

 \be L_{eff}^{(i)} =- \sum^\infty_{n=2} \frac{N_{top}}{2n N_c^2V_4}
 \prod^n_{i=1} \int\frac{d^4 p_i}{(2\pi)^4} T (p_1, p_2, ... p_n)
 \varphi^{(t_1)}_{LR} (p_1) \varphi^{(t_1)}_{RL} (p_2) ... \varphi^{(t_n)}_{RL} (p_n)
 \label{A12}\ee
 with the notation
 \be T(p_1,... p_n)\equiv \prod^{n-1}_{i=1,3,...} \lambda (q_i,
 q_{i+1});\lambda (q_i,
 q_{i+1})={4\rho^4 k(q_i) k(q_{i+1}) (q_iq_{i+1})}
 \label{A13}\ee
 and $ q_i\equiv p_i-p_{i+1}$ with $p_{n+1} \equiv p_1$, and  $q_\mu k(q)$ is the Fourier transform of the instanton field, $$
 k(q) \equiv \frac{1}{q^2}\left [K_2 (q\rho) -\frac{2}{(q\rho)^2}\right ].$$

 Comparing (\ref{A12}) with eqs. (\ref{3.2}), (\ref{3.5}) from
 \cite{5}, one can define the correlators $J_n$,
 \be J_n (p_1, ... p_n)=\frac{N}{V_42n  N_c^2}  T (p_1, p_2,
 ... p_n).\label{A14}\ee

 Using the asymptotics of the modified Bessel function $K_2(z)$ one has
 $$ q^2k(q) \sim -\frac12 - O((q\rho)^2), ~~ (q\rho)\to 0,$$
 \be q^2k(q) \sim -\frac{2}{(q\rho)^2}+ \sqrt{\frac{\pi}{2q\rho}}e^{-q\rho}\left (1+O\left (\frac{1}{q\rho}\right)\right ), ~~ (q\rho)\to
 \infty.\label{A15}\ee
 Hence the equation for $\mu(p)$, eq. (\ref{3.5}) of \cite{5}
 acquires the form (with $\varphi_{LR} = id^+, \varphi_{RL} =id)$
 \be
 \mu(p) = \frac{N}{4V_4N^2_c } \sum^\infty_{n=2,4..} \int \frac{ d^4 p_1...
 d^4 p_{n-1}(-)^{ \frac{n}{2}}}{(2\pi)^{4(n-1)}} T(p, p_1, ..., p_{n-1})
  d^+ (p_1) d (p_2) d^+(p_3)... d(p_{n-2}) d^+(p_{n-1}), \label{A16}\ee
 with $d(p) \equiv \frac{\mu(p)}{p^2+\mu^2(p)}$,  $d^+(p) =\frac{\mu^+(p)}{p^2+(\mu^+(p))^2}$ and  equation for $\mu^+(p)$
 is obtained by replacement $\mu\leftrightarrow\mu^+, d\leftrightarrow d^+$.
 The integral
 in  (\ref{A16}) converges both at small and large momenta, in the
 latter case symbolically as $ \frac{N}{2V_4 q^3} \left(
 \frac{\mu(p) d^4p}{q^3(p^2+\mu^2(p))}\right)^{n-1}.$

 As a next step we consider the contribution of instanton zero
 modes to the quark propagator in the random instanton field, i.e.
 we consider as in \cite{5} the total averaged field of randomized
 instantons $C_\mu(x)$, which produces higher field correlators
 $J_n(x_1, ... x_n)$, and in addition a number of topcharges
 (possibly also instantons), which produce zero modes. As in Eq.
 (\ref{3.12}) of \cite{5} one can write

 \be C_\mu(x) \to \bar C_\mu(x) + \sum^N_{i=1} A_\mu^{(i)} (x-
 R_i).\label{A17}\ee

 Note, that in the randomized ensemble $\bar C_\mu(x)$ we are not
 interested by the zero modes, while in the second term in
 (\ref{A17}) only zero modes effects will be taken into account.
 Correspondingly, the fermion propagator assumes the form (cf. Eq.
 (\ref{3.17}) of \cite{5}, where difference in $\mu, \mu^+$ was not taken into account)
 \be S(p) =\frac{1}{\hat p- i \bar \mu (p)} +
 \frac{\bar c_0(p) l_+}{-i\mu_0},~~ S^{-1} = \frac{(p^2+\mu\mu^+)}{p^2+\mu  q} (\hat p -iql_--i\mu l_+)\label{A18}\ee
 where
 $$ l_\pm = \frac{1\pm \gamma_5}{2},~~ q\equiv \mu +  \frac{(p^2+\mu\mu^+ )\bar c_0}{\mu_0},$$
  \be \bar \mu(p) = \frac{\mu(p) + \mu^+(p)}{2} + \gamma_5
 \frac{\mu(p)-\mu^+(p)}{2}\label{19}\ee
 and the term $\mu(p)$ comes from $\Phi_{LR}$, while $\mu^+(p)$--
 from $\Phi_{RL}$. Here $\bar c_0(p)$ and $\mu_0$ are obtained from
 zero modes of the ensemble of topcharges of the second term on
 the r.h.s. of (\ref{A17}). For simplicity we shall consider
 instead one-instanton zero mode in the eigenvalue expansion ($R$
 is the position of instanton)
 \be S(x,y) = \sum_n \frac{u_n(x-R) u_n^+ (y-R)}{\lambda_n -i \mu}
 .\label{A20}\ee

 In case, when the effective fermion mass $\mu=\mu(x,y)$ depends
 on coordinates,  (\ref{A20}) transforms into

 \be S(x,y) = \sum_n \frac{u_n(x-R) u^+_m (y-R)}{\lambda_{n}
 \delta_{nm} - i \mu_{nm}}, ~~ \mu_{nm} = \lan n | \mu(x,y) |
 m\ran, \label{A21}\ee

 One can separate in (\ref{A21}) the zero mode term and associate
 $\mu_0$ in (\ref{A18}) with $\mu_{00}$
 \be \mu_0 \equiv  \mu_{00} = \int\mu (p) u^+_0 (p) u_0 (p)
 \frac{d^4 p}{(2\pi)^4}\label{A22}\ee
while nonzero modes are assembled in the first term in (\ref{A17}). As was
discussed in \cite{5}, the zero mode coefficient is proportional to $|u_0
(p)|^2$, and here we in addition to Eq. (\ref{3.18}) of \cite{5}, also notice
that one instanton produces a zero mode of definite  chirality, i.e. \be c_0
(p) = \frac{N_0}{V_4} | u_0 (p)|^2 (1+\gamma_5)\equiv \bar c_0(p)
\frac{1+\gamma_5}{2}\label{A23}\ee where we generalized to the case of $N_0$
instantons in the volume $V_4$.

We turn now to the zero mode wave function. The spacial part is given by \be
u_0(p) =\frac{\rho}{\pi} \int \frac{d^4xe^{ipx}}{(\rho^2 +x^2)^{3/2}} = 2 \rho
\int^\infty_0 \frac{J_0 (pr) r^3 dr}{(\rho^2+ r^2)^{3/2}}.\label{A24}\ee

The last integral can be expressed via generalized hypergeometric series
$$~_1F_2(a; b, c|z) = \left(1+\frac{a}{b\cdot c} \cdot \frac{z}{1!} +
\frac{a(a+1)}{b(b+1)c(c+1)} \frac{z^2}{2!}+...\right)$$ \be u_0 (p) = 2 \rho
\left\{ \frac{1}{p}~_1F_2 \left(\frac32; \frac12, \frac12,
\frac{p^2p^2}{4}\right) - 2 \rho ~_1F_2 \left( 2; \frac32, 1;
\frac{p^2p^2}{4}\right)\right\}.\label{A25}\ee

Expansion at small $p$ is

\be u_0 (p) = 2 \rho \left\{ \frac{1}{p}- 2 \rho + \frac32 p\rho^2
+O(p^2)\right\}.\label{A26}\ee

Calculation of $d(p)$ using (\ref{A18})  with $c_0(p)$ from (\ref{A23}) yields
$$ d(p) = \frac{\delta tr \ln S^{-1}(p)}{\delta\mu(p)}=
 \frac{ \mu^+(p)}{2(p^2+ \mu(p)\mu^+(p))(1+\frac{\bar c_0(p)}{\mu_0} \mu(p))}+
$$

\be
 +\frac{u^2_0 (p)}{2\mu_0^2} \int \frac{d^4p'}{(2\pi)^4}\frac{\mu(p')\bar c_0(p')}{(1+\frac{\bar c_0(p')}{\mu_0} \mu(p'))}
 \label{A27}\ee

\be d^+(p) =\frac{\delta tr \ln S^{-1}(p)}{\delta\mu^+(p)}= \frac{ \mu (p)\left
[1-\frac{p^2\bar c_0(p)}{\mu\mu_0 (1+\frac{\bar c_0}{\mu_0}\mu)}\right
]}{2(p^2+\mu\mu^+)}\label{A28}\ee

One can see, that $d(p) \neq d^+(p)$, and the extra term in (\ref{A27}), as
compared to (\ref{A28}), contains additional factor $\frac{1}{\mu_0^2}$.

To the lowest order in $c_0\left(c_0 \sim
\left(\frac{\rho}{R}\right)^4\right)$, one has \be
d(p)\cong\frac{u^2_0(p)}{2\mu^2_0}\int \frac{d^4p'}{(2\pi)^4} \mu (p') \bar c_0
(p') + \frac{\mu^+(p)}{2(p^2+\mu\mu^+)} \label{A30}\ee \be d^+(p) =\frac12
\frac{ \mu (p)}{p^2+\mu \mu^+(p)}.\label{A31}\ee

One can see, that the small values of $\mu_0$ and $\mu(p)$ can occur as
solutions of (\ref{A16}) due to the first  term in (\ref{A30}).

Note, that the first  term on the r.h.s. of  (\ref{A30}) corresponds to the
right zero mode $\sim \frac{1+\gamma_5}{2}$, and $\mu_0$ in the denominator of
this term contains $\mu(p)$. In the opposite case, when $c_0(p) \sim
\frac{1-\gamma_5}{2}, ~\mu_0$ is proportional to $\mu^+(p)$ and, consequently,
$d(p)$ and $d^+(p)$ in (\ref{A28}), (\ref{A30}) interchange their places.

\vspace{2cm}
 \setcounter{equation}{0}
\renewcommand{\theequation}{A3.\arabic{equation}}

{\bf \large
\noindent Appendix  3}\\

\noindent{\bf \large Derivation of equations (\ref{33}) for the mass matrix  }\\

In the case of  one flavor (or diagonal flavor matrix) the terms $\varphi_{LR}
= \varphi_1^+= id^+, \varphi_{RL} \equiv \varphi_1=id$ were  found in Appendix
2 and can be written as (in case of no zero modes) \be d(p) =
\frac{\mu^+(p)}{2(p^2+\mu\mu^+)}, ~~ d^+ (p) =\frac12
\frac{\mu(p)}{(p^2+\mu\mu^+)}\label{A3.1}\ee and  more general form see in
(\ref{A30}), (\ref{A31}).

We now consider the case, when $\{\mu_i\}, \{\mu^+_i\}$, when $\{\varphi_i\},
\{\varphi_i\}, i=1,2,3,4$ form $SU(2)$ matrices in the flavor space. In this
case \be G=\frac{1}{\partial^2+m^2}= \frac{1}{\hat a+\hat b\gamma_5}, ~~ \hat a
= \partial^2 + \frac{\mu^2+\mu^{+2}}{2}; ~~ \hat b =
\frac{\mu^2-\mu^{+2}}{2}\label{A3.2}\ee and the dash signs over $\hat a, \hat
b$ denote matrices in flavor space. In case, when $\mu$ and $\mu^+$ are
matrices with equal elements $b\equiv 0$ and one returns to expressions
(\ref{A3.1}). However, when $\mu\neq \mu^+$, $G$ can be written as \be
G=(1-\gamma_5 \hat a^{-1}\hat b) \frac{1}{1-(\hat a^{-1}\hat b)^2}\hat
a^{-1},\label{A3.3}\ee and the pole structure of $G$ is more complicated.

In (\ref{35}) we have used (\ref{A3.1}), in case when  valid for $\mu=\mu^+$,
and in this case $K^+_{ik}= \bar K^+_{ik} =0, $ and $\varphi_i
=\varphi_i^+=\frac{\mu_i(p)}{2(p^2+\hat\mu^2)}$. \\ In this case  $\hat m^2
\equiv \hat \mu^2 = \left( \begin{array}{ll} \mu^2_1+\mu_3\mu_4, &
(\mu_1+\mu_2) \mu_3\\
(\mu_1+\mu_2)\mu_4,& \mu^2_2+\mu_3\mu_4)\end{array}\right)$.  Inserting this
expression into (\ref{29}), one reproduces the system of equations (\ref{35}).

\vspace{2cm}
 \setcounter{equation}{0}
\renewcommand{\theequation}{A4.\arabic{equation}}

{\bf \large
\noindent Appendix  4}\\

\noindent{\bf \large }\\

We calculate here the term $tr (G\Omega G\Omega) $, which we can write as
follows: \be - tr G\Omega G\Omega = tr (\hat \partial  \hat V +\hat V \hat
\partial+ i\hat V^2) G (\hat \partial \hat V + \hat V \hat \partial G + i\hat
V^2).\label{A1.1}\ee

Writing explicitly  in the  coordinate space, one has \be (\hat \partial \hat V
+\hat V \hat \partial ) G = \int \gamma_\mu \gamma_\nu \left (
\frac{\partial}{\partial x_\mu} V_\nu  (x,y) + V_\mu (x,y)
\frac{\partial}{\partial y_\nu}\right) G(y,z) d^4y.\label{A1.2}\ee In the last
term on the r.h.s. of (\ref{A1.2}), one can differentiate by parts yielding \be
V_\mu (x,y) \frac{\partial}{\partial y_\nu} G(y,z) dy =\int
\frac{\partial}{\partial  y_\nu}( V_\mu (x,y) G (y,z)) d^4 y- \int
(\frac{\partial}{\partial u_\nu} V_\mu (x,y)) G (y,z) dy.\label{A1.3}\ee Here
the first term vanishes, and the second yields the minus sign, which allows to
rewrite \be (\hat \partial \hat V+ \hat V \hat \partial) G \to (\hat \partial
\hat V - \hat V \hat{\bar\partial}) G\label{A1.4}\ee and for the total answer
of terms with derivatives one obtains \be - tr G\Omega G\Omega =4(\partial_\mu
V_\nu)^2 -4 \partial_\mu V_\nu \partial_\nu V_\mu+...\label{A1.5}\ee where
$V_\mu \equiv V^A_{1\mu} t^A + V_{2\mu} y(a)$, so that finally one has \be - tr
G\Omega G\Omega   = ((F^A_{\mu\nu})^2+ 2 F^2_{\mu\nu} \sum^4_{a=1} y^2(a))G^2+
\left( \frac12 (V^A_{1\mu})^2+ V^2_{2\mu} y^2\right) G^2,\label{A1.6}\ee where
$$ F^A_{\mu\nu} =\partial_\mu V^A_{1\nu} -\partial_\nu V^A_{1\mu} + e_{ABC}
V^B_{1\mu} V^C_{1\nu},~~ F_{\mu\nu} =\partial_\mu V_{2\nu} -
\partial_\nu V_{2\mu}$$ and $G_uG_d\approx G^2_u\approx
G^2_d\equiv G^2$,
and
 $$ tr (G\hat \Omega G\hat \Omega)
= - (F^A_{\mu\nu})^2 G_u G_d  - F^2_{\mu\nu} 2\sum_a y^2(a)+$$\be+
\left(\frac12 (V^A_\mu)^2+ V^2_{2\mu}y^2\right)^2G^2 +...,\label{3.17}\ee

 which enters in the
renormalization factor $Z_W$, $\sqrt{Z_W} V_{1\mu}^A=\tilde V^A_{1\mu}$\be
\frac{N_2}{4} (F^A_{\mu\nu})^2 G^2 \equiv \frac{Z_W}{4}
(F^A_{\mu\nu})^2=\frac14 (\tilde F^A_{\mu\nu})^2\label{3.19}\ee and the same
for $F_{\mu\nu}, V_{2\mu}$. Note, that after renormalization the renormalized
$(V)^4$ term enters with coefficient $1/Z$, $(V)^4=\frac{\tilde V^4}{Z},$ where
$Z\simeq \frac{N_2}{16\pi^2}\ln \frac{M^2}{\mu^2}$ and hence for large $M$ this
term can be neglected, $N_2=4$ for $SU(4)$ group.

The calculations done above in agreement with \cite{*,**} support the idea,
that the quark loop expansion yields an effective Lagrangian for the composite
vector field, where the gauge invariance is restored (seen e.g. in appearance
of $F^A_{\mu\nu}$ and $F_{\mu\nu}$ in (\ref{3.17})), up to the mass terms.

Let us now return to our case, where in addition to the term $V_L^At^A$ there
is also $V_R^At^A$ and diagonal in $SU(2)$ indices terms $V_L$ and $V_R$. The
EWSB combination is now \be tr (\hat G\hat N\hat G\hat N) = tr \{ (\bar V_L
\hat\mu-\hat \mu\bar V_R) \hat G  (\bar V_L \hat \mu-\hat\mu\bar V_R) \hat
G\}.\label{42}\ee

To simplify matter, one can keep first only $\hat V_L$, then the general
structure of (\ref{42}) comes out to be \be (42)= \bar c_1(V_L^A)^2 + (\bar c_3
V_L^A n^A+ \bar c_2 V_L)^2 + \bar c_4 (V_L^An^A)^2\label{43}\ee where $\bar
c_i$ depend on $\mu_i$ and $\hat G$. One can see in (\ref{43}) explicitly the
EWSB, which manifests itself both in the mixing between terms $V_L^A$ and
$V_L$, and in the last term on the r.h.s. of (\ref{43}). Note, that  the vector
$n^A$ defines the ``direction of the diagonalization'' of the matrix $\hat
\mu$, $\hat \mu=(a+\ven\vetau)  \mu_d(a^*-i\ven\vetau)$, which is the $3d$ axis
in the case  of Standard Model. Thus one can visualize the effects of EWSB,
however, one also obtains extra terms (e.g. the last term in (\ref{43})) which
should be missing in the realistic case.   In the case, when also $V^a_R$ and
$V_R$ are present, one can have spontaneous violation of left-right symmetry,
however the analysis in the general case, when both $\bar V_l$ and $\bar V_R$
are nonzero and $\mu\neq \mu^+$ becomes rather complicated and will be reported
elsewhere \cite{22}.


\begin{thebibliography}{99}


\bibitem{1} PDG: K.Nakamura et al., JPG  {\bf  37}, 075021 (2010).

\bibitem{2} G.Cveti\v{c}, Rev. Mod. Phys. {\bf 71}, 513 (1999),
hep-ph/9702381.

\bibitem{3} K.Lane,  arXiv:hep-ph/0202255.

\bibitem{4}
C.T.Hill and E.H.Simmons, Phys. Rep. {\bf 381}, 235 (2003); Erratum-ibid. {\bf
390},
  553 (2004).

\bibitem{4*}
S. Weinberg, Phys. Rev. D {\bf 13}, 974 (1976);\\
 L. Susskind, Phys. Rev. D {\bf 20},
2619 (1979);\\
 B. Holdom, Phys. Lett. B {\bf 150}, 301 (1985);\\
  K. Yamawaki, M.
Bando, and K.-i. Matumoto, Phys. Rev. Lett. {\bf 56}, 1335 (1986);\\
 T. W.
Appelquist, D. Karabali, and L. C. R. Wijewardhana, Phys. Rev. Lett. {\bf  57},
957 (1986).

\bibitem{5*} F.Sannino, Acta Phys. Polon. B{\bf 40}, 3533 (2009), arXiv:
0911.0931.


\bibitem{5} Yu.A.Simonov, Yad.Fiz {\bf 73}, 1944 (2010),   0912.1946 [hep-ph].

\bibitem{6} Yu.A.Simonov,  Phys. Lett. Nucl. (in press), 1004.2672 [hep-ph].

\bibitem{7}  J.C.Pati, A.Salam, Phys. Rev. Lett. {\bf 31}, 661 (1973), Phys. Rev.
{\bf D 10}, 275 (1974); R.N.Mahapatra and J.C.Pati, Phys. Rev. D {\bf 11}, 566,
2558 (1975).

\bibitem{8*} J.C.Pati, Phys. Rev.  D {\bf 68}, 072002; hep-ph/0209160; Proc. of
the XI Int. workshop ``Neutrino telescopes'', Venice, 2005; hep-ph/0507307.

\bibitem{8}  Yu.A.Simonov, ``Confinement'', Phys. Usp. {\bf 39}, 313 (1996),
arXiv: hep-ph/9709344.

\bibitem{11} C.G.Callan, R.Dashen and D.J.Gross,  Phys.
Rev.  D{\bf17}, 2717 (1978), ibid D {\bf 19}, 1826 (1987);\\ E.V.Shuryak, Nucl.
Phys. B {\bf 203}, 93, 116, 140 (1982);\\D.I.Diakonov and V.Yu.Petrov, Nucl.
Phys. B {\bf 245}, 259 (1984).

\bibitem{9} Yu.A.Simonov, Phys. Atom. Nucl. {\bf  60}, 2069 (1997), arXiv: hep-ph/ 9704301.


\bibitem{*} M.Veltman, Acta Phys. Pol. B{\bf 12}, 437 (1981);\\
S.Mandelstam, A.Passion in Physics, edited by C. De Tar et al., (World
scientific, Singapore, 1985), p. 97.
\bibitem{**} M.Suzuki, arXiv:1006.1319; Phys. Rev. D {\bf 37}, 210 (1988).
\bibitem{***}A.Cohen, H.Georgi and E.Simmons, Phys. Rev, {\bf 38},
405 (1988).
\bibitem{10} A.V.Nefediev, Yu.A,Simonov, M.A.Trusov,  Int.J.Mod. Phys. E {\bf
18}, 549 (2009).



\bibitem{12} P.V.Pobylitsa, Phys.
  Lett. B {\bf 226},  387 (1989).
  \bibitem{13} Yu.A.Simonov, Phys. Lett.  B {\bf 412}, 371 (1997).

  \bibitem {14}
  H.G.Dosch, Phys. Lett. B {\bf 190},  177 (1987);\\
  H.G.Dosch and Yu.A.Simonov, Phys. Lett. B {\bf 205},  339 (1988);\\
   Yu.A.Simonov, Nucl.  Phys.  B {\bf 307},  512 (1988);\\
 A.Di Giacomo, H.G.Dosch, V.I.Shevchenko, Yu.A.Simonov, Phys. Rept. {\bf  372},  319
 (2002).


\bibitem{15}

A.Yu.Dubin, A.B.Kaidalov, and Yu.A.Simonov, Phys. Lett.  B {\bf 323},  41
(1994).

  \bibitem{16}
 Yu.A.Simonov, Ann. Phys. {\bf 323}, 783 (2008).
 \bibitem{17} A.A.Belavin, A.M.Polyakov, A.S.Schwartz and Yu.S.Tyupkin,   Phys. Lett. B {\bf 59}, 85 (1975).

 \bibitem{18} M.Creutz, ``Quark,  Gluons and Lattices'', Cambridge Univ. Press, Cambridge, 1983.

\bibitem{24}V.A.Miransky, M.Tanabashi, and K.Yamawaki, Phys. Lett. B{\bf 221},
177 (1989); Mod. Phys. Lett. A {\bf 4}, 1043 (1989).


\bibitem{22}Yu.A.Simonov (in preparation).

\bibitem{26}M.Hashimoto and V.A.Miransky, Phys. Rev. D {\bf 80}, 013004 (2009),
[arXiv: 0901.4354].

\bibitem{27}G.Senjanovic and R.N.Mohapatra, Phys. rev. D   {\bf 12}, 1502 (1975).
\bibitem{28} D.Guadagnoli and R.N.Mohapatra, arXiv:1008.1074.

\bibitem{29} A.J.Buras, K.Gemmler and G.Isidori, Nucl. Phys. B {\bf 843}, 107
(20011), [arXiv: 1007.1993, hep-ph].

\bibitem{30} H.Georgi, in {\it Particles and Fields}, ed. C.Carlson (AIP, New
York, 1975);\\H.Fritzsch and P.Minkowski, Ann. Phys. {\bf93}, 193 (1975).

\bibitem{31} C.H.Albright and S.M.Barr, Phys. Rev. D {\bf 62}, 093008 (2000),
hep-ph/0210374;\\C.S.Aulakh, B.Bajc, A.Melfo, G.Senjanovic, and F.Vissani,
Phys. Lett. B {\bf 588} , 196 (2004),  hep-ph/0306242.

\bibitem{32} S.Bertolini, L.Di Luzio and M.Malinsk\'{y}, arXiv:1010.0338,
[hep-ph].
 \end{thebibliography}
\end{document}